\documentclass[fleqn,twoside]{article}
\usepackage{espcrc2}

\usepackage{graphicx}
\usepackage[figuresright]{rotating}

\newcommand{\AmS}{{\protect\the\textfont2
  A\kern-.1667em\lower.5ex\hbox{M}\kern-.125emS}}

\def\lnA{{\langle\ln A\rangle}}
\def\th{$^{\rm th}$}

\title{Dissecting the knee --- Air shower measurements with KASCADE}

\author{
 J.R.~H{\"o}randel\address[1]
               {Institut f\"ur Experimentelle Kernphysik, University
              of Karlsruhe, 76021~Karlsruhe, Germany\\[-2mm]}%
	      \thanks{http://www-ik.fzk.de/$\sim$joerg.},
 {T.~Antoni}\addressmark[1],
 {W.D.~Apel}\address[2]
         {Institut f\"ur Kernphysik, Forschungszentrum Karlsruhe, 
          76021~Karlsruhe, Germany\\[-2mm]},
 {A.F.~Badea}\addressmark[2]\thanks
                          {on leave of absence from NIPNE, Bucharest, Romania},
 {K.~Bekk}\addressmark[2],
 {A.~Bercuci}\addressmark[2]\footnotemark[2],
 {H.~Bl\"umer}\addressmark[1]\addressmark[2],
 {H.~Bozdog}\addressmark[2],
 {I.M.~Brancus}\address[3] 
         {National Institute of Physics and Nuclear Engineering,
              7690~Bucharest, Romania\\[-2mm]},
 {C.~B\"uttner}\addressmark[1],
 {A.~Chilingarian}\address[4]
         {Cosmic Ray Division, Yerevan Physics Institute,
              Yerevan~36, Armenia\\[-2mm]},
 {K.~Daumiller}\addressmark[1],
 {P.~Doll}\addressmark[2],
 {R.~Engel}\addressmark[2],
 {J.~Engler}\addressmark[2],
 {F.~Fe{\ss}ler}\addressmark[2],
 {H.J.~Gils}\addressmark[2],
 {R.~Glasstetter}\addressmark[2]\thanks
            {now at Universit\"at Wuppertal, 42119 Wuppertal, Germany},
 {A.~Haungs}\addressmark[2],
 {D.~Heck}\addressmark[2],
 {K-H.~Kampert}\addressmark[1]\addressmark[2]\footnotemark[3],
 {H.O.~Klages}\addressmark[2],
 {G.~Maier}\addressmark[2],
 {H.J.~Mathes}\addressmark[2],
 {H.J.~Mayer}\addressmark[2],
 {J.~Milke}\addressmark[2],
 {M.~M\"uller}\addressmark[2],
 {R.~Obenland}\addressmark[2],
 {J.~Oehlschl\"ager}\addressmark[2],
 {S.~Ostapchenko}\addressmark[1]\thanks
            {on leave of abscence from Moscow State University, Moscow, Russia},
 {M.~Petcu}\addressmark[3],
 {H.~Rebel}\addressmark[2],
 {A.~Risse}\address[5]
          {Soltan Institute for Nuclear Studies, 90950~Lodz, Poland\\[-2mm]},
 {M.~Risse}\addressmark[2],
 {M.~Roth}\addressmark[1],\\
 {G.~Schatz}\addressmark[2],
 {H.~Schieler}\addressmark[2],
 {J.~Scholz}\addressmark[2],
 {T.~Thouw}\addressmark[2],
 {H.~Ulrich}\addressmark[2],
 {J.~van~Buren}\addressmark[2],
 {A.~Vardanyan}\addressmark[4],
 {A.~Weindl}\addressmark[2],
 {J.~Wochele}\addressmark[2], and
 {J.~Zabierowski}\addressmark[5]
 }
       
\begin{document}
\begin{abstract}
Recent results of the KASCADE air shower experiment are presented in order to
shed some light on the astrophysics of cosmic rays in the region of the knee in
the energy spectrum. The results include investigations of high-energy
interactions in the atmosphere, the analysis of the arrival directions of
cosmic rays, the determination of the mean logarithmic mass, and the unfolding
of energy spectra for elemental groups.
\vspace*{-1mm}
\end{abstract}

\maketitle

\section{Introduction}
The origin of high-energy cosmic rays is among the most interesting questions
in astrophysics.  The origin of a structure in the all-particle energy spectrum
around 4~PeV, the so-called knee, is generally believed to be a corner stone in
the understanding of the astrophysics of high-energy cosmic rays.  The knee is
proposed to be caused by the maximum energy reached in cosmic-ray accelerators
or due to leakage of particles from the Galaxy. Hence, an understanding of the
origin of the knee reveals hints on the acceleration and propagation of cosmic
rays.  Experimental access to the understanding of the sources, acceleration
and propagation mechanisms is provided by detailed investigation of the arrival
directions, energy spectra, and mass composition of the ultrarelativistic
particles.  

While at energies below 1~PeV cosmic rays can be measured directly
at the top of the atmosphere, the strongly decreasing flux as
function of energy requires large acceptances and exposure times
for higher energies. Presently they can be realized in ground based
facilities only. There, the secondary products, generated by
interactions of high-energy cosmic-ray particles in the atmosphere,
the extensive air showers, are registered. It turns out that a
correct description of the high-energy interactions in the
atmosphere is crucial for a precise astrophysical interpretation of
air shower measurements.

To investigate cosmic rays from several $10^{13}$ eV up to beyond $10^{17}$~eV
the air shower experiment KASCADE ("Karlsruhe Shower Core and Array DEtector")
\cite{kascade} is operated since 1996. The experiment detects the three main
components of air showers simultaneously.  A $200\times 200$~m$^2$ scintillator
array measures the electromagnetic and muonic components.  The 320~m$^2$
central detector system combines a large hadron calorimeter with several muon
detection systems.  In addition, high-energy muons are measured by an
128~m$^2$ underground muon tracking detector.

\section{High-energy interactions in the atmosphere}\label{wwtest}
A correct understanding of high-energy interactions in the atmosphere is
indispensable for a good astrophysical interpretation of air shower data.  The
electromagnetic part of the showers is well understood and described by QED.
For the air shower development the understanding of multi-particle production
in hadronic interactions with a small momentum transfer is essential. Due to
the energy dependence of the coupling constant $\alpha_s$ and the resulting
large values for soft interactions, the latter cannot be calculated within QCD
using perturbation theory.  Instead, phenomenological approaches have been
introduced in different models.

For the numerical simulation of the development of air showers the program
CORSIKA \cite{corsika} is widely used.  It offers the possibility to use
different models to describe low and high-energy hadronic interactions. A
principal objective of the KASCADE experiment is to investigate the air shower
development in detail and test the validity of the models included in
simulation codes such as CORSIKA, using as much information as possible from
the simultaneous measurement of the electromagnetic, muonic and hadronic
components.  

With these investigations already several problems in existing codes could be
pointed out and some interaction models (or particular versions of them) could
be shown to be incompatible with the measured data
\cite{wwtest,milke,milkeisvh,jrhwwtest}.
\begin{figure}\centering
  \includegraphics[width=\columnwidth]{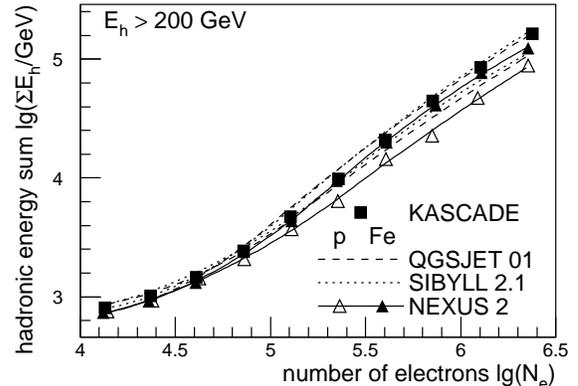}
  \vspace*{-15mm}
  \caption{\small Hadronic energy sum versus number of electrons.\vspace*{-5mm}}
  \label{ehne}
\end{figure}
As an example of present activities, Fig.~\ref{ehne} shows the hadronic energy
sum as function of the number of electrons. Presented are measured values
compared to predictions of three different interaction models for showers
induced by primary protons or iron nuclei.  For a presentation of the data as
function of the number of electrons one expects an enrichment of light
primaries within the particular intervals, hence, the data should approach the
values for the proton component.  One recognizes that the data are compatible
with the predictions of QGSJET and SIBYLL, while on the other hand NEXUS~2
predicts too less hadronic energy in most of the electron number range.  From
such distributions one can conclude that the present version of NEXUS is not
compatible with the data.  More detailed investigations of the models QGSJET~01
and SIBYLL~2.1 are presently in progress.

\section{Arrival directions} \label{arrival}
The investigation of the arrival directions of cosmic rays improves the
understanding of the propagation of the particles through the Galaxy and their
sources.  Model calculations of the diffusion process in the Galactic magnetic
field indicate that there could be an anisotropy on a scale of $10^{-4}$ to
$10^{-2}$ depending on particle rigidity as well as strength and structure
of the Galactic magnetic field \cite{candia}.  Diffusion models relate a
rigidity dependent leakage of particles from the Galaxy to the steepening (or
knee) in the all-particle energy spectrum around 4~PeV. Thus, anisotropy
measurements can provide substantial information on the origin of the knee.

In KASCADE investigations \cite{maier,schatz} attention has been drawn to a
search for point sources and large-scale anisotropy.  Of special interest is
the search for potential gamma-ray induced showers. Since photons are not
deflected in the Galactic magnetic field, they are suitable for a direct search
for the sources of high-energy particles. Experimentally such an investigation 
is realized by the selection of muon poor showers.

\begin{figure}\centering
\vspace*{-3mm}
\includegraphics[width=0.9\columnwidth]{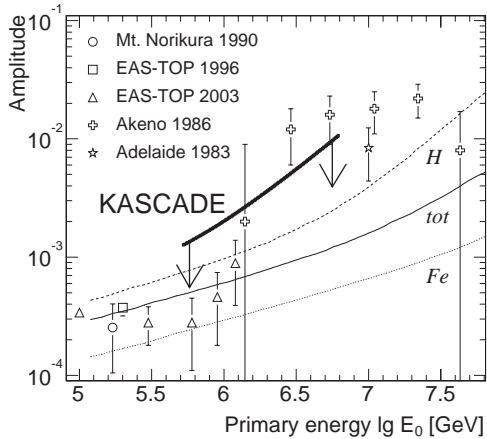}
\vspace*{-10mm}
\caption{\small Rayleigh amplitudes $A$ as function of primary energy. The 
	 KASCADE upper limit (bold line) is compared to results from the
	 literature \cite{nagashima,aglietta,kifune,gerhard}. Model predictions
	 \cite{candia} for the total anisotropy as well as for the light and
	 heavy component are also shown (thin lines).\vspace*{-4mm}}
\label{aniso}
\end{figure}

A search for point sources was performed for primary photon candidates as well
as for all (charged) particles.  The search covers the whole sky, visible by
KASCADE (declination $15^\circ$ to $80^\circ$).  Special attention was drawn to
the Galactic plane and known gamma-ray sources in the TeV region.  The search
was complemented by an analysis of the most energetic showers registered and an
investigation of the most photon-like primary particles.  None of the searches
reveals an indication for a significant excess of the flux.  The distributions
of the significance values over the whole sky follow the expectations for an
isotropic cosmic-ray flux.

The results for the analysis of the large-scale anisotropy are illustrated in
Fig.\ref{aniso}. There, the Rayleigh amplitude is plotted versus the primary
energy. The upper limit derived from the KASCADE data is compatible with other
measurements from the literature and also with theoretical calculations
\cite{candia} for different elemental groups.

\section{Energy spectra for elemental groups} \label{espek}
\begin{figure*} \centering
\includegraphics[width=0.99\columnwidth]{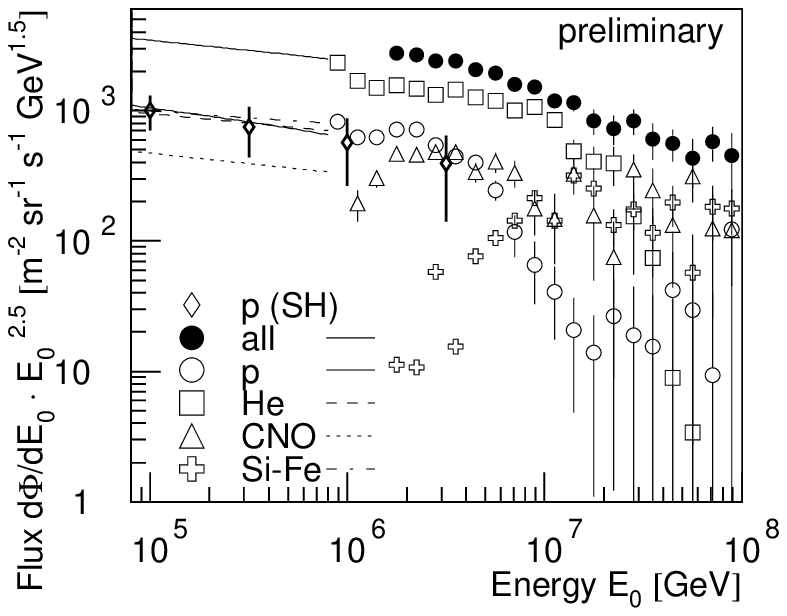}
\includegraphics[width=0.99\columnwidth]{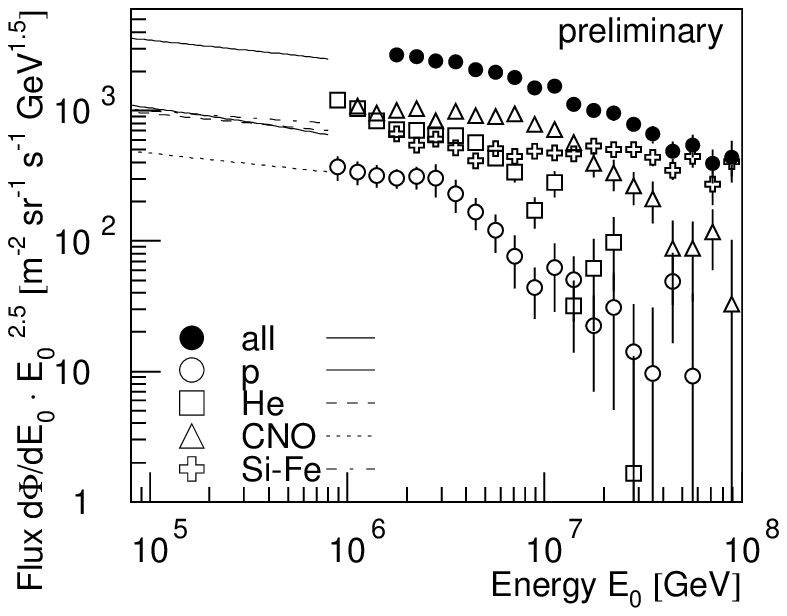}
\vspace*{-10mm}
\caption{\small Energy spectra for groups of elements derived from the data
	 using CORSIKA with the hadronic interaction models QGSJET (left) and
	 SIBYLL (right). The lines indicate extrapolations of direct
	 measurements \cite{polygonato}.\vspace*{-3mm}}
\label{espec}
\end{figure*}
The most direct experimental access to the astrophysics of cosmic rays is
provided by measurements of their energy spectrum and mass composition.
Investigations of the KASCADE experiment reveal that the knee in the
all-particle spectrum is caused by a turn-off of the light component (protons
and helium nuclei) \cite{muden}. This is implies an increase of the mean
logarithmic mass ($\lnA$) as function of energy.  While it is beyond doubt that
the data reveal such an increase, the absolute values of $\lnA$ depend on the
observables investigated and on the interaction models used in the simulations
to interpret the data \cite{jrhlna,rothnn,buettner}. The differences amount to
about $\Delta \lnA\approx1$. This indicates that further and more detailed
investigations of high-energy interactions in the atmosphere are necessary for
an unambiguous astrophysical interpretation of the observed data.

Presently, the most promising approach is the unfolding of energy spectra for
individual elemental groups from the data of the electromagnetic and muonic
component \cite{ulrich,ulrichisvh,rothisvh,roth}.  Systematic studies are
performed with different hadronic interaction models available in CORSIKA, for
both, low and high-energy ($>80$~GeV) interactions. Different unfolding methods
are applied, in order to investigate the systematic effects introduced by the
individual methods. The different methods result in similar flux values if the
same code is used for the air shower simulation.  But significant differences
occur for different interaction models.  This is illustrated in
Fig.~\ref{espec}, where recent results of an analysis applying the Gold
algorithm are depicted.

The figure also shows the flux of primary protons as obtained in an analysis of
unaccompanied hadrons \cite{mueller}.  The flux is compatible with the proton
flux as obtained from the unfolding procedure.  For comparison, also the
results of direct measurements at lower energies at the top of the atmosphere
\cite{polygonato} are given in the figure.

The analyses indicate that, at present, the understanding of primary cosmic
rays is limited by the insufficient knowledge of high-energy hadronic
interactions in the atmosphere and not by too low statistics of the
measurements or the systematic uncertainties introduced by different
reconstruction methods.

\section{Conclusion and outlook}
The present status of the KASCADE experiment to investigate high-energy cosmic
rays in the knee region has been discussed.  Continuing efforts are taken in
order to improve the understanding of high-energy interactions in the
atmosphere.  A search for point sources of charged particles as well as gamma
rays did not reveal any significant excess. Upper limits for the large scale
anisotropy have been derived.  The observations reveal an increase of the mean
logarithmic mass as function of energy in the knee region.  Energy spectra for
groups of elements have been derived.  To extend the investigations to higher
energies up to $5\times10^{18}$~eV data taking with an enlarged array
\cite{grande} has started in 2003.

{\small
\footnotetext[1]{KASCADE Collaboration}

}

\end{document}